\documentclass[twocolumn,english,superscriptaddress,showpacs,pra,aps]{revtex4}
\usepackage[T1]{fontenc}
\usepackage[latin9]{inputenc}
\usepackage{bm}
\usepackage{amsmath}
\usepackage{amssymb}
\usepackage{graphicx}
\usepackage{esint}
\usepackage{CJK}
\usepackage{babel}

\begin{document}

\title{Non-unital non-Markovianity of quantum dynamics}

\author{Jing Liu}
\affiliation{Zhejiang Institute of Modern Physics, Department of Physics, Zhejiang
University, Hangzhou 310027, China}

\author{Xiao-Ming Lu}
\email{luxiaoming@gmail.com}
\affiliation{Centre for Quantum Technologies, National University of Singapore,
3 Science Drive 2, Singapore 117543, Singapore}

\author{Xiaoguang Wang}
\email{xgwang@zimp.zju.edu.cn}
\affiliation{Zhejiang Institute of Modern Physics, Department of Physics, Zhejiang
University, Hangzhou 310027, China}
\begin{abstract}
Trace distance is available to capture the dynamical information of
the unital aspect of a quantum process. However, it cannot reflect
the non-unital part. So, the non-divisibility originated from the
non-unital aspect cannot be revealed by the corresponding measure
based on the trace distance. We provide a measure of non-unital
non-Markovianity of quantum processes, which is a supplement to
Breuer-Laine-Piilo (BLP) non-Markovianity measure. A measure on the
degree of the non-unitality is also provided.
\end{abstract}

\pacs{03.65.Yz, 03.67.-a, 03.65.Ta}

\maketitle

\section{Introduction}

\begin{figure}[bp]
\centering{}\includegraphics[width=6cm]{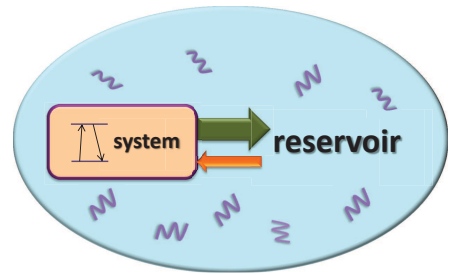}\caption{\label{fig:sketch}(Color online)~Sketch of the information flow
picture for non-Markovianity~\cite{Breuer2009}. According to this
scenario, the loss of distinguishability of the system's states indicates
the information flow from the system to the reservoir. If the dynamics
is Markovian, the information flow is always outward, represented
by the green thick arrow. Non-Markovian behaviors occurs when there
is inward information flow, represented by the orange thin arrow,
bringing some distinguishability back to the system.}
\end{figure}

Understanding and characterizing general features of the dynamics
of open quantum systems is of great importance to physics, chemistry,
and biology~\cite{breuer}. The non-Markovian character is one of
the most central aspects of an open quantum process, and attracts increasing
attentions~\cite{Wolf2008,Wolf2008-1,Breuer2009,Rivas2010,Lu2010,Chruscinski2010,Zhang2012,Hou2011,Luo2012,Vasile2011,Laine2012,Liu2011,Bylicka2013,Laine2010,Rajagopal2010}.
Markovian dynamics of quantum systems is described by a quantum dynamical
semigroup~\cite{breuer,Lendi_book}, and often taken as an approximation
of realistic circumstances with some very strict assumptions. Meanwhile,
exact master equations, which describe the non-Markovian dynamics,
are complicated~\cite{Zhang2012}. Based on the infinitesimal divisibility
in terms of quantum dynamical semigroup, Wolf \textit{et al.} provided
a model-independent way to study the non-Markovian features~\cite{Wolf2008-1,Wolf2008}.
Later, in the intuitive picture of the backward information flow leading
to the increasing of distinguishability in intermediate dynamical
maps, Breuer, Laine, and Piilo (BLP) proposed a measure on the degree
of non-Markovian behavior based on the monotonicity of the trace distance
under quantum channels~\cite{Breuer2009}, as shown in  Fig.~\ref{fig:sketch}.
The BLP non-Markovianity has been widely studied, and applied in various
models~\cite{Xu2010,Chruscinski2011,Rebentrost2011,Wibmann2012,Haikka2011,Clos2012}.

Unlike for classical stochastic processes, the non-Markovian criteria
for quantum processes is non-unique, and even controversial. First,
the non-Markovian criteria from the infinitesimal divisibility and
the backward information flow are not equivalent~\cite{Haikka2011,Chruscinski2011}.
Second, several other non-Markovianity measures, based on different
mechanism like the monotonicity of correlations under local quantum
channels, have been introduced~\cite{Rivas2010,Luo2012}. Third,
even in the framework of backward information flow, trace distance
is not the unique monotone distance for the distinguishability between
quantum states. Other monotone distances on the space of density operators
can be found in Ref.~\cite{Petz1996}, and the statistical distance~\cite{Wootters1981,Braunstein1994}
is another widely-used one. Different distance should not be expected
to give the same non-Markovian criteria. The inconsistency among various
non-Markovianity reflects different dynamical properties.

In this paper, we show that the BLP non-Markovianity cannot reveal
the infinitesimal non-divisibility of quantum processes caused by
the non-unital part of the dynamics. Besides non-Markovianity, ``non-unitality''
is another important dynamical property, which is the necessity for
the increasing of the purity $\mathrm{Tr\rho^{2}}$ under quantum
channels~\cite{Lidar2006} and for the creating of quantum discord
in two-qubit systems under local quantum channels~\cite{Streltsov2011}.
In the same spirit as BLP non-Markovianity, we define a measure on
the non-unitality. As BLP non-Markovianity is the most widely used
measure on non-Markovianity, we also provide a measure on the non-unital
non-Markovianity, which can be conveniently used as a supplement to
the BLP measure, when the quantum process is non-unital. We also give
an example to demonstrate an extreme case, where the BLP non-Markovianity
vanishes while the quantum process is not infinitesimal divisible.

This paper is organized as follows. In Sec.~\ref{Review}, we give
a brief review on the representation of density operators and quantum
channels with Hermitian orthonormal operator basis, and various measures
on non-Markovianity. In Sec.~\ref{sec:Non-unital-NM}, we investigate
the non-unitality and the non-unital non-Markovianity and give the
corresponding quantitative measures respectively. In Sec.~\ref{sec:EXAMPLE},
we apply the non-unital non-Markovianity measure on a family of quantum
processes, which are constructed from the generalized amplitude damping
channels. Section \ref{sec:Conclusion} is the conclusion.

\section{Review on quantum channels and non-Markovianity\label{Review}}

\subsection{Density operators and quantum channels represented by Hermitian operator
basis.}

The states of a quantum system can be described by the density operator
$\rho$, which is positive semidefinite and of trace one. Quantum
channels, or quantum operations, are completely positive and trace-preserving
(CPT) maps from density operators to density operators, and can be
represented by Kraus operators, Choi-Jamio\l{}kowski matrices, or
transfer matrices~\cite{Wolf_channels,Kraus1983,Choi1975,Jamiolkowski1972}.

In this work, we use the Hermitian operator basis to express operators
and represent quantum channels. Let $\{\lambda_{\mu}\mid\mu=0,1,\cdots,d^{2}-1\}$
be a complete set of Hermitian and orthonormal operators on complex space $\mathbb{C}^{d}$,
i.e., $\lambda_{\mu}$ satisfies $\lambda_{\mu}^{\dagger}=\lambda_{\mu}$
and $\langle\lambda_{\mu},\lambda_{\nu}\rangle:=\mathrm{Tr}(\lambda_{\mu}^{\dagger}\lambda_{\nu})=\delta_{\mu\nu}$.
Any operator $O$ on $\mathbb{C}^{d}$ can be express by a column
vector $r:=(r_{0},r_{1},\cdots,r_{d^{2}-1})^{\mathrm{T}}$ through
\begin{equation}
O=\sum_{\mu=0}^{d^{2}-1}r_{\mu}(O)\lambda_{\mu}
\end{equation}
with $r_{\mu}(O):=\langle\lambda_{\mu},O\rangle$. Every $r_{\mu}(O)$
is real if $O$ is Hermitian.

In the meantime, any quantum channel
$\mathcal{E}\colon\rho\mapsto\mathcal{E}(\rho)$ can be represented
by $T(\mathcal{E})\colon r(\rho)\mapsto r[\mathcal{E}(\rho)]$ via
\begin{equation}
r[\mathcal{E}(\rho)]=T(\mathcal{E})r(\rho) \label{eq:1},
\end{equation}
where $T(\mathcal{E})$ is a $d^2\times d^2$ real matrix with the elements
\begin{equation}
T_{\mu\nu}(\mathcal{E}):=\left\langle \lambda_{\mu},\mathcal{E}(\lambda_{\nu})\right\rangle. \label{eq:T}
\end{equation}
Furthermore, one can easily check that
\begin{equation}
T(\mathcal{E}_{1}\circ\mathcal{E}_{2})
=T(\mathcal{E}_{1})T(\mathcal{E}_{2})
\end{equation}
for the composition of quantum channels. Here $\mathcal{E}_{1}\circ\mathcal{E}_{2}$
denotes the composite maps $\mathcal{E}_{1}(\mathcal{E}_{2}(\rho))$.

Taking into the normalization of the quantum states, i.e., $\mathrm{Tr}(\rho)=1$,
$r_{0}$ can be fixed as $r_{0}(\rho)=1/\sqrt{d}$ for any density
operator $\rho$ by choosing $\lambda_{0}=\openone/\sqrt{d}$ with
$\openone$ the identity operator. In such a case, $\lambda_{\mu}$
for $\mu=1,2,\cdots,d^{2}-1$ are traceless and generate the algebra
$\mathfrak{su}(d)$. This real parametrization $r_{\mu}(\rho)$ for
density operators is also called as coherent vector, or generalized
Bloch vector~\cite{Bloch1946,Hioe1981,Bengtsson2006}. In order to eliminate
the degree of freedom for the fixed $r_{0}$, we use the decomposition
$r=(r_{0},\mathbf{r})^{\mathrm{T}}$. Therefore, any density operator $\rho$ can be
expressed as
\begin{equation}
\rho=\frac{\openone}{d}+\mathbf{r}\cdot\bm{\lambda},\label{eq:rho}
\end{equation}
with $\mathbf{r}$
the generalized Bloch vector and $\bm{\lambda}$ represents $(\lambda_{1},\lambda_{2},\cdots,\lambda_{d^{2}-1})^{\mathrm{T}}$.
Under this frame, quantum channels can be represented by the affine
map~\cite{Lendi_book,Nielsen}
\begin{equation}
\mathbf{r}(\mathcal{E}(\rho))=M(\mathcal{E})\mathbf{r}(\rho)+\mathbf{c}
(\mathcal{E}) \label{eq:2},
\end{equation}
where $M(\mathcal{E})$ is a real matrix with the dimension $d^2-1$ and the elements of the vector $\mathbf{c}(\mathcal{E})$ reads
\begin{equation}
\left[\mathbf{c}(\mathcal{E})\right]_{\mu} = \left\langle\lambda_{\mu},\mathcal{E}(\openone)\right\rangle/d,
\end{equation}
for $\mu=1,2,\cdots,d^{2}-1$. Comparing Eq.~(\ref{eq:1}) with Eq.~(\ref{eq:2}), one could find that
\begin{equation}
T_{\mu\nu}(\mathcal{E})=[M(\mathcal{E})]_{\mu\nu},
\end{equation}
for $\mu,\nu=1,2,\cdots,d^{2}-1$.
Thus, $T(\mathcal{E})$ can be decomposed into the following sub-blocks:
\begin{equation}
T(\mathcal{E})=\left[\begin{array}{c|c}
1 & 0_{1\times(d^{2}-1)}\\
\hline \sqrt{d}\mathbf{c} & M
\end{array}\right].
\end{equation}

Reminding that a quantum channel $\mathcal{E}$ is said to be unital
if and only if $\mathcal{E}(\openone/d)=\openone/d$~\cite{Nielsen}, one could find that the necessary and sufficient condition for a unital map is that $\mathbf{c}(\mathcal{E})=0$, namely,
\begin{equation}
\mathbf{c}(\mathcal{E})=0\Longleftrightarrow\mathcal{E}\text{ is unital}.
\end{equation}
Thus, $\mathbf{c}(\mathcal{E})$ describes the non-unital property of
the quantum channel $\mathcal{E}$. The necessary and sufficient condition above could be easily proved by realizing that the Bloch vector of $\openone/d$ is zero vector, i.e., $\mathbf{r}=0$.
Based on the sub-block form of $T(\mathcal{E})$, $\mathbf{c}(\mathcal{E})=0$ is equivalent to that $T(\mathcal{E})$ is block diagonal, i.e., $T(\mathcal{E})=\mathrm{diag}(1,M(\mathcal{E}))$.

Whether a quantum channel $\mathcal{E}$ is completely positive (CP) can be reflected by the Choi-Jamio\l{}kowski matrix~\cite{Choi1975,Jamiolkowski1972}
\begin{equation}
C(\mathcal{E}):=\left(\mathcal{E}\otimes\openone\right)
(|\Omega\rangle\langle\Omega|), \label{eq:C}
\end{equation}
where $|\Omega\rangle=\frac{1}{\sqrt{d}}\sum_{j=0}^{d-1}|j\rangle \otimes|j\rangle$ is the maximally entangled state. Here $\{|j\rangle\} $ is a basis in Hilbert space. $\mathcal{E}$ is
CP if and only if the Choi-Jamio\l{}kowski matrix is positive.
With the Hermitian operator basis, $|\Omega\rangle\langle\Omega|$ is
a $d^2\times d^2$ matrix and can be written in the form~\cite{Yu2009}
\begin{equation}
|\Omega\rangle\langle\Omega|=\frac{1}{d}\sum_{\nu=0}^{d^{2}-1}\lambda_{\nu}\otimes\lambda_{\nu}^{\mathrm{T}}.
\end{equation}
Substituting this formula into Eq.~(\ref{eq:C}) and utilizing Eq.~(\ref{eq:T}), one could express the Choi-Jamio\l{}kowski matrix as
\begin{equation}
C(\mathcal{E})=\sum_{\mu,\nu=0}^{d^{2}-1}\frac{1}{d}T_{\mu\nu}
(\mathcal{E})\lambda_{\mu}\otimes\lambda_{\nu}^{\mathrm{T}}.
\end{equation}
If $\mathcal{E}$ is unital, it can be reduced into
\begin{equation}
C(\mathcal{E})=\frac{1}{d^2}\left(\openone_{d^2\times d^2}+d\sum_{\mu,\nu=1}^{d^2-1} M_{\mu\nu}\lambda_{\mu} \otimes \lambda^{\mathrm{T}}_{\nu}\right).
\end{equation}

\subsection{Non-divisibility and non-Markovianity}

Without the presence of correlation between the open system and its
environment in the initial states, the reduced dynamics for the open
system from $t=0$ to any $t\geq0$ can be expressed as
\begin{equation}
\mathcal{E}_{t,0}:\rho\mapsto\mathrm{Tr}_{E}\left[U(t)\left(\rho\otimes\rho_{E}\right)U(t)^{\dagger}\right],
\end{equation}
which is a quantum channel. This indicates that $\mathcal{E}_{t,0}$ is CPT. The unitary operator $U(t)$ describes the time evolution of the closed entirety, and $\rho_{E}$ is the
initial state of the environment. A quantum process $\mathcal{E}_{t}:=\mathcal{E}_{t,0}$
is said to be infinitesimal divisible, also called as time-inhomogeneous
or time-dependent Markovian, if it satisfies the following composition
law \cite{Wolf2008}
\begin{equation}
\mathcal{E}_{t_{2},0}=\mathcal{E}_{t_{2},t_{1}}\circ\mathcal{E}_{t_{1},0}
\end{equation}
for any $t_{2}\ge t_{1}\geq0$, where $\mathcal{E}_{t_{2},t_{1}}$
is also completely positive and trace preserving.

Various measures on the degree of the non-Markovian behavior of quantum
processes have been proposed and investigated~\cite{Breuer2009,Rivas2010,Vasile2011,Luo2012,Hou2011}.
Almost all of the measures on the non-Markovianity can be classified
into three kinds, base on the degree of the violation of the following
properties owned by the infinitesimal divisible quantum process:

(i) Monotonicity of distance $D$ under CPT maps. That is $D(\mathcal{E}(\rho_{1}),\mathcal{E}(\rho_{2}))\leq D(\rho_{1},\rho_{2})$
for any quantum channel $\mathcal{E}$, where $D(\rho_{1},\rho_{2})$
is an appropriate monotone distance under CPT maps on the space of density operators~\cite{Petz1996}, including trace distance, Bures distance, statistical distance, relative entropy, and fidelity (although fidelity itself is not a distance,
it can be used to construct monotone distances) and so on. Some measures
on non-Markovianity by increasing of the monotone distance during
the mediate dynamical maps $\mathcal{E}_{t_{2},t_{1}}$ have been
given and discussed in Refs.~\cite{Breuer2009,Vasile2011}.

The typical measure of this type, which would be used later in this paper, was first proposed by Breuer, Laine, and Piilo in Ref.~\cite{Breuer2009}, based on the monotonicity of trace distance \cite{Nielsen,Bengtsson2006}
\begin{equation}
D_{\mathrm{tr}}(\rho_{1},\rho_{2}):=\frac{1}{2}\mathrm{Tr}\left|\rho_{1}-\rho_{2}\right|,
\end{equation}
where $|O|:=\sqrt{O^{\dagger}O}$.
Interpreting the increase of the trace distance during the time evolution as the information flows from the environment back to the system, the definition of the BLP non-Markovianity is defined by
\begin{equation}
\mathcal{N}_{\mathrm{BLP}}(\mathcal{E}_{t}):=\max_{\rho_{1},\,\rho_{2}}\int_{\sigma>0}\mathrm{d}t\,\sigma\left(t,\rho_{1},\rho_{2}\right),
\end{equation}
where
\begin{equation}
\sigma\left(t,\rho_{1},\rho_{2}\right):=\frac{\mathrm{d}}{\mathrm{d}t}D_{\mathrm{tr}}\left(\rho_{1}(t),\rho_{2}(t)\right),
\end{equation}
and $\rho_{i}(t)=\mathcal{E}_{t}(\rho_{i})$ for $i=1,2$ are two
evolving states.

(ii) Positivity of the Choi-Jamio\l{}kowski matrix for CPT maps. The
Choi-Jamio\l{}kowski matrix $C(\mathcal{E})\geq0$ if and only if
$\mathcal{E}$ is a quantum channel, namely, $\mathcal{E}$ is a CPT
map. Some measures on non-Markovianity by the negativity of the Choi-Jamio\l{}kowski
matrix for mediate dynamical maps $\mathcal{E}_{t_{2},t_{1}}$have
been given and discussed in Refs.~\cite{Rivas2010,Hou2011}.

In this work we would use one of these measures, which was proposed by
Rivas, Huelga and Plenio (RHP) in Ref.~\cite{Rivas2010}. They utilize the negativity of the Choi-Jamio\l{}kowski matrix $C$ for the mediate dynamical maps with the definition
\begin{equation}
\mathcal{N}_{\mathrm{RHP}}(\mathcal{E}_{t}):=\int_{0}^{\infty}g(t)dt,
\end{equation}
where
\begin{equation}
g(t):=\underset{\epsilon\rightarrow0^{+}}{\mathrm{lim}}
\frac{\mathrm{Tr}|C(\mathcal{E}_{t+\epsilon,t})|-1}{\epsilon}.
\label{eq:gt}
\end{equation}

(iii) Monotonicity of correlations $E$ under local quantum channels.
That is $E\left[\left(\mathcal{E}\otimes\openone\right)(\rho^{AB})\right]\leq E(\rho^{AB})$
for any local quantum channel $\mathcal{E}$, where $E$ is an appropriate
measure for the correlations in the bipartite states $\rho^{AB}$,
including entanglement entropy and the mutual information. The corresponding
measures on non-Markovianity are given and discussed in Refs.~\cite{Rivas2010,Luo2012}.

\section{Non-unital non-Markovianity\label{sec:Non-unital-NM}}

The non-Markovianity measure $\mathcal{N}_{\mathrm{BLP}}$ is available to capture the non-Markovian behavior of the unital aspect of the dynamics. But for the non-unital
aspect, it is not capable. To show this, we use the Hermitian
orthonormal operator basis to express states and quantum channels.
Utilizing Eq.~(\ref{eq:rho}), the trace distance between two states $\rho_{1}$ and $\rho_{2}$
is given by
\begin{eqnarray}
D_{\mathrm{tr}}\left(\rho_{1},\rho_{2}\right) & = & \frac{1}{2}\mathrm{Tr}\big|\left[\mathbf{r}(\rho_{1})-\mathbf{r}(\rho_{2})\right]\cdot\bm{\lambda}\big|.
\end{eqnarray}
Therefore, for the two evolving states, we get
\begin{equation}
D_{\mathrm{tr}}\left(\rho_{1}(t),\rho_{2}(t)\right)=\frac{1}{2}\mathrm{Tr}
\big|M(\mathcal{E}_{t})\left[\mathbf{r}(\rho_{1})-\mathbf{r}(\rho_{2})\right]
\cdot\bm{\lambda}\big|, \label{eq:dis}
\end{equation}
where $\rho_{1}$, $\rho_{2}$ are initial states of the system.

From this equation one can see that the trace distance between any
two evolved states is irrelevant to the non-unital part $\mathbf{c}(\mathcal{E}_{t})$
of the time evolution. Then if there are two quantum channels, whose
affine maps are $\mathbf{r}\mapsto M\mathbf{r}+\mathbf{c}_{1}$ and
$\mathbf{r}\mapsto M\mathbf{r}+\mathbf{c}_{2}$, respectively,
the characteristic of trace distance between the evolving states from
any two initial states cannot distinguish these two channels. More
importantly, $\mathbf{c}(\mathcal{E}_{t})$ may cause the non-divisibility
of the quantum process $\mathcal{E}_{t}$, and this cannot be revealed
by $\mathcal{N}_{\mathrm{BLP}}$.

On the other hand, the non-unital part $\mathbf{c}(\mathcal{E}_{t})$
has its own physical meaning: $\mathbf{c}(\mathcal{E}_{t})\neq0$
is necessary for the increasing of the purity $\mathcal{P}(\rho)=\mathrm{Tr}(\rho^{2})$~\cite{Lidar2006}.
In other words,
\begin{equation}
\mathbf{c}(\mathcal{E}_{t})=0\ \Longrightarrow\ \mathcal{P}(\mathcal{E}_{t}(\rho))\leq\mathcal{P}(\rho),\ \forall\rho.
\end{equation}
Besides the non-Markovian feature, the non-unitality is another kind
of general feature of quantum processes. In analogy to the definition
of BLP non-Markovianity, we defined the following measure on the degree
of the non-unitality of a quantum process:
\begin{equation}
N_{\mathrm{nu}}(\mathcal{E}_{t}) = \max_{\rho_{0}}
\int_{\frac{\mathrm{d}}{\mathrm{d}t}\mathcal{P}(\mathcal{E}_{t}(\rho_{0}))>0}
\left| \frac{\mathrm{d} \mathcal{P}[\mathcal{E}_{t}(\rho_{0})] }{\mathrm{d}t} \right|
\mathrm{d}t,
\end{equation}
where $\rho_{0}$ is the initial state. Obviously, $N_{\mathrm{nu}}(\mathcal{E}_{t})$ vanishes if $\mathbf{c}(\mathcal{E}_{t})=0$.

Since the non-unital aspect of the dynamics, which is not revealed by the trace distance, has its own speciality, we aim to measure the effect of non-unitality on non-Markovian behavior.
However, a perfect separation of the non-unital aspect from the
total non-Markovianity may be infeasible.  Therefore we require a weak version
$\mathcal{N}_\mathrm{nu}$ for measuring non-unital non-Markovianity to satisfy the following three conditions:
(i) $\mathcal{N}_{\text{nu}}$ vanishes if $\mathcal{E}_{t}$ is infinitesimal divisible,
(ii) $\mathcal{N}_{\text{nu}}$ vanishes if $\mathcal{E}_{t}$ is unital,
(iii) $\mathcal{N}_{\mathrm{nu}}$ should be relevant to $\mathbf{c}(\mathcal{E}_{t})$. Based on these
conditions, we introduce the following measure
\begin{equation}
\mathcal{N}_{\mathrm{nu}}:=\underset{\varrho_{\tau}\in\mathcal{X}}{\max}\int\limits _{\sigma_{\text{nu}}>0}\sigma_{\text{nu}}(t,\,\varrho_{\tau})\text{d}t,\label{eq:definition_NUN}
\end{equation}
where $\mathcal{X}:=\left\{ \varrho_{\tau}\mid0\leq\tau\leq\infty\right\} $
with $\varrho_{\tau}:=\mathcal{E}_{\tau}(\openone/d)$ is the
set of the trajectory states which evolve from the maximally mixed
state, and
\begin{equation}
\sigma_{\text{nu}}(t,\varrho_{\tau}):=\frac{\text{d}}{\text{d}t}D\left[\mathcal{E}_{t}(\varrho_{0}),
\mathcal{E}_{t}(\varrho_{\tau})\right],
\end{equation}
with $D(\rho_{1},\rho_{2})$ an appropriate distance which will be
discussed below. The first condition is guaranteed if we require that
$D$ is monotone under any CPT maps, i.e., $D[\mathcal{E}(\rho_{1}),\mathcal{E}(\rho_{2})]\leq D(\rho_{1},\rho_{2})$
for any quantum channel $\mathcal{E}$. For the unital time evolution,
the set $\mathcal{X}=\{\openone/d\}$ only contains the maximally
mixed state, so the above defined $\mathcal{N}_{\mathrm{nu}}$ vanishes,
and the second condition is satisfied. The third condition excludes
the trace distance.

In this paper, we use the Bures distance which is defined as
\begin{equation}
D_{\mathrm{B}}(\rho_{1},\rho_{2})=\sqrt{2\left[1-F\left(\rho_{1},\rho_{2}\right)\right]},
\end{equation}
where
\begin{equation}
F(\rho_{1},\rho_{2})=\mathrm{Tr}|\sqrt{\rho_{1}}\sqrt{\rho_{2}}|
=\mathrm{Tr}\sqrt{\sqrt{\rho_{1}}\rho_{2}\sqrt{\rho_{1}}}
\end{equation}
is the Uhlmann fidelity~\cite{Uhlmann1976,Jozsa1994} between $\rho_{1}$
and $\rho_{2}$. Here $|O|=\sqrt{O^{\dagger}O}$.
Bures distance is an appropriate distance for $\mathcal{N}_{\mathrm{nu}}$ because it
obeys the monotonicity under CPT maps~\cite{Petz1996} and is relevant to $\mathbf{c}(\mathcal{E}_{t})$.
As here only the monotonicity of distance is relevant, for simplicity,
we can also take the square of the Bures distance or just the opposite value of Unlmann
fidelity as a simple version of monotone ``distance''~\cite{Vasile2011}.
Quantum relative entropy~\cite{Vedral2002}
$S(\rho_{1}\Vert\rho_{2})=\text{Tr}\left[\rho_{1}(\ln\rho_{1}-\ln\rho_{2})\right]$,
or its symmetric version
$S_{\mathrm{sym}}(\rho_{1}\Vert\rho_{2}):=S(\rho_{1}\Vert\rho_{2})+S(\rho_{2}\Vert\rho_{1})$,
is another qualified candidate for the distance.
Noting that when the support of $\rho_{1}$ is not within the support
of $\rho_{2}$, namely, $\mathrm{supp}(\rho_{1})\nsubseteq\mathrm{supp(\rho_{2})}$,
$S(\rho_{1}\Vert\rho_{2})$ will be infinite, so in such cases, quantum
relative entropy will bring singularity to the measure of non-Markovianity.
Also, Hellinger distance~\cite{Luo2004} is qualified. Although all of these
distances are monotone under CPT maps, they may have different characteristics
in the same dynamics, see Ref.~\cite{Dajka2011}.

The difference between non-unital non-Markovian measure defined by
Eq.~(\ref{eq:definition_NUN}) and the BLP-type measures, including those which use other alternative distances, is the restriction on the pairs of initial states. Comparing with the BLP-type measures relying on any pair of initial states, the non-unital non-Markovianity measure only relies on the pairs consisting of the maximally mixed state and its trajectory states. On one hand, this restriction makes the non-unital non-Markovianity measure vanish when the quantum processes are unital, no matter they are Markovian or non-Markovian; on the other hand, this restriction reflects that non-unital non-Markovianity measure reveals only a part of information concerning the non-Markovian behaviors.

\section{EXAMPLE \label{sec:EXAMPLE}}

To illustrate the non-unital non-Markovian behavior, we give an example
in this section. We use the generalized amplitude damping channel (GADC) as
a prototype to construct a quantum process. The GADC can be described
by $\mathcal{E}(\rho)=\sum_{i}E_{i}\rho E_{i}^{\dagger}$ with the
Kraus operators $\{E_{i}\}$ given by~\cite{Fujiwara2004,Nielsen}
\begin{eqnarray}
E_{1} & = & \sqrt{p}\left(\begin{array}{cc}
1 & 0\\
0 & \sqrt{\eta}
\end{array}\right),\\
E_{2} & = & \sqrt{p}\left(\begin{array}{cc}
0 & \sqrt{1-\eta}\\
0 & 0
\end{array}\right),\\
E_{3} & = & \sqrt{1-p}\left(\begin{array}{cc}
\sqrt{\eta} & 0\\
0 & 1
\end{array}\right),\\
E_{4} & = & \sqrt{1-p}\left(\begin{array}{cc}
0 & 0\\
\sqrt{1-\eta} & 0
\end{array}\right),
\end{eqnarray}
where $p$ and $\eta$ are real parameters. Note that for any $p\in[0,1]$
and any $\eta\in[0,1]$, the corresponding $\mathcal{E}$ is a quantum
channel. For a two-level system, the Hermitian orthonormal operator
basis can be chosen as $\bm{\lambda}=\bm{\sigma}/\sqrt{2}$, where $\bm{\sigma}=\{\sigma_{x},\sigma_{y},\sigma_{z}\}$ is
the vector of Pauli matrices. With the decomposition in Eq.~(\ref{eq:rho}),
the affine map for the Bloch vector is given by $\mathbf{r}(\mathcal{E}(\rho))\mapsto M(\mathcal{E})\mathbf{r}(\rho)+\mathbf{c}(\mathcal{E})$~\cite{Nielsen},
where
\begin{eqnarray}
M(\mathcal{E}) & = & \left(\begin{array}{ccc}
\sqrt{\eta} & 0 & 0\\
0 & \sqrt{\eta} & 0\\
0 & 0 & \eta
\end{array}\right),\\
\mathbf{c}(\mathcal{E}) & = & \left(0,0,\frac{(2p-1)(1-\eta)}{\sqrt{2}}\right)^{\mathrm{T}}.
\end{eqnarray}
The GADC is unital if and only if $p=1/2$ or $\eta=1$. When $\eta=1$,
$M(\mathcal{E})=\openone$, the map is identity.

A quantum process can be constructed by making the parameter $p$
and $\eta$ to be dependent on time $t$. For simplicity, we take
$p_{t}=\cos^{2}\omega t$ and $\eta_{t}=e^{-t}$, where $\omega$
is a constant real number. This is a legitimate quantum process, because
$\mathcal{E}_{t}$ is a quantum channel for every $t\geq0$, and $\mathcal{E}_{t=0}$
is the identity map.

First, let us consider the $\mathcal{N}_{\mathrm{BLP}}$ for this
quantum process. For any two initial states $\rho_{1}$ and $\rho_{2}$,
we have the trace distance
\begin{align}
D_{\mathrm{tr}}\left[\mathcal{E}_{t}(\rho_{1}),\mathcal{E}_{t}(\rho_{2})\right] & =\frac{1}{2}\mathrm{Tr}\left|M(\mathcal{E}_{t})[\mathbf{r}(\rho_{1})-\mathbf{r}(\rho_{2})]\cdot\frac{\bm{\sigma}}{\sqrt{2}}\right|\nonumber \\
 & =\frac{1}{\sqrt{2}}\left|M(\mathcal{E}_{t})[\mathbf{r}(\rho_{1})-\mathbf{r}(\rho_{2})]\right|,
\end{align}
where $|\mathbf{r}|=\sqrt{\mathbf{r}\cdot\mathbf{r}}$ is the Euclidean
length of the vector $\mathbf{r}$, and we used the equality
\begin{equation}
(\mathbf{a}\cdot\bm{\sigma})(\mathbf{b}\cdot\bm{\sigma})=(\mathbf{a}\cdot\mathbf{b})\openone
+i\bm{\sigma}\cdot(\mathbf{a}\times\mathbf{b})
\end{equation}
for Pauli matrices. Denoting $\mathbf{r}(\rho_{1})-\mathbf{r}(\rho_{2})$
by $(x,y,z)^{\mathrm{T}}$, we get
\begin{equation}
D_{\mathrm{tr}}\left[\mathcal{E}_{t}(\rho_{1}),\mathcal{E}_{t}(\rho_{2})\right]=\frac{e^{-t/2}}{\sqrt{2}}\sqrt{x^{2}+y^{2}+e^{-t}z^{2}},\label{eq:GADC_tr_distance}
\end{equation}
which implies $\frac{\mathrm{d}}{\mathrm{d}t}D_{\mathrm{tr}}\left[\mathcal{E}_{t}(\rho_{1}),\mathcal{E}_{t}(\rho_{2})\right]\leq0$
for every time point $t\geq0$ and for any real numbers $x$, $y$,
and $z$. Thus, the BLP non-Markovianity vanishes, i.e., $\mathcal{N}_{\mathrm{BLP}}(\mathcal{E}_{t})\equiv0$,
although $\mathcal{E}_{t}$ may be not infinitesimal divisible, which
will become clear later.

\begin{figure}
  \begin{centering}
    \includegraphics[width=5cm]{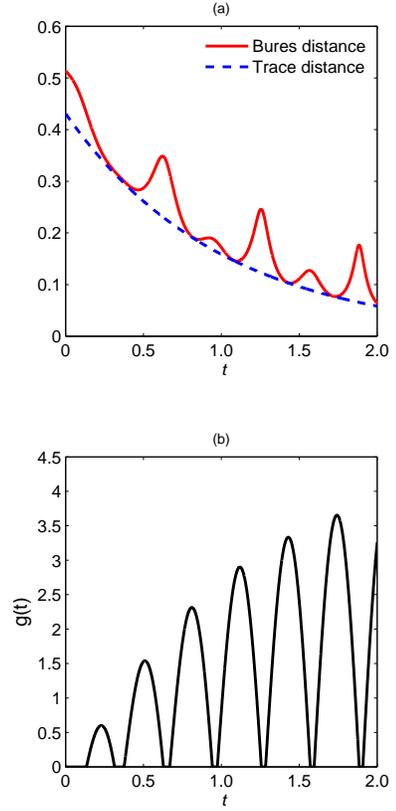}
  \end{centering}
  \caption{\label{fig:DB_Dtr}
    (Color online)~(a)~Evolution of trace distance and Bures distance between
    two evolving states of a two-level system under the variant generalized
    amplitude damping channel, initially from the maximal mixed states
    $\varrho_{0}=\openone/2$ and its trajectory state
    $\varrho_{\tau}=\mathcal{E}_{\tau}(\varrho_{0})$, respectively.
    (b)~The evolution of $g(t)$ defined by Eq.~(\ref{eq:gt}), whose
    integral with respect to time $t$ is RHP measure for non-Markovianity. In these figures, the
    parameters are taken as $\tau=10$ and $\omega=5$.}
\end{figure}

In order to investigate whether $\mathcal{E}_{t}$ is infinitesimal
divisible or not, we shall apply $\mathcal{N}_{\mathrm{nu}}$ in the
above model. The trajectory of the maximally mixed state under $\mathcal{E}_{t}$
reads
\begin{equation}
\mathcal{E}_{t}(\varrho_{0})=\frac{1}{2}\openone+\mathbf{c}_{t}\cdot\frac{\bm{\sigma}}{\sqrt{2}}=\frac{1}{2}\left(\begin{array}{cc}
1+W_{t} & 0\\
0 & 1-W_{t}
\end{array}\right),
\end{equation}
where
\begin{equation}
W_{t}:=(2p_{t}-1)(1-\eta_{t})=\cos(2\omega t)(1-e^{-t}).
\end{equation}
Taking these trajectory states as the initial states, we get the corresponding
evolving states:
\begin{eqnarray}
\mathcal{E}_{t}(\varrho_{\tau}) & = & \frac{1}{2}\openone+(M_{t}\mathbf{c}_{\tau}+\mathbf{c}_{t})\cdot\frac{\bm{\sigma}}{\sqrt{2}}\\
 & = & \frac{1}{2}\left(\begin{array}{cc}
1+W_{t}+\eta_{t}W_{\tau} & 0\\
0 & 1-W_{t}-\eta_{t}W_{\tau}
\end{array}\right).\quad
\end{eqnarray}
Then the fidelity reads
\begin{equation}
F\left[\mathcal{E}_{t}(\varrho_{0}),\mathcal{E}_{t}(\varrho_{\tau})\right]=\frac{1}{2}(h_{+}+h_{-}),
\end{equation}
where
\begin{eqnarray}
h_{+} & := & \sqrt{\left(1+W_{t}\right)\left(1+W_{t}+\eta_{t}W_{\tau}\right)},\\
h_{-} & := & \sqrt{\left(1-W_{t}\right)\left(1-W_{t}-\eta_{t}W_{\tau}\right)}.
\end{eqnarray}
To compare with the behavior of trace distance, we also get $D_{\mathrm{tr}}\left[\mathcal{E}_{t}(\varrho_{0}),\mathcal{E}_{t}
(\varrho_{\tau})\right]=\left|\eta_{t}W_{\tau}\right|/2$.
With the expressions $\eta_{t}=e^{-t}$ and $p_{t}=\cos^{2}\omega t$,
it is
\begin{equation}
D_{\mathrm{tr}}\left[\mathcal{E}_{t}(\varrho_{0}),\mathcal{E}_{t}(\varrho_{\tau})\right]=\frac{e^{-t}}{2}\left|\cos2\omega\tau\right|(1-e^{-\tau}).\label{eq:GADC_nonunital}
\end{equation}

In Fig.~\ref{fig:DB_Dtr}(a), we can see that while the trace distance
between the evolving states $\mathcal{E}_{t}(\varrho_{0})$ and
$\mathcal{E}_{t}(\varrho_{\tau})$ monotonously decreases with
the time $t$, the Bures distance increases during some intermediate
time intervals. From Eq.~(\ref{eq:GADC_nonunital}), one can see although
$D_{\mathrm{tr}}[\mathcal{E}_{t}(\varrho_{0}),\mathcal{E}_{t}(\varrho_{\tau})]$
depends on $W_{\tau}$, it does not depend on $W_{t}$. Actually,
from Eq.~(\ref{eq:GADC_tr_distance}) one could find that for any
two initial states, the trace distance between the evolving states
is independent on $W_{t}$. In this sense, the BLP non-Markovianity
treats a family of quantum processes, which only differ with $p_{t}$,
as the same one. Meanwhile, $\mathcal{N}_{\mathrm{nu}}$ reveals the
effects of $p_{t}$ on the infinitesimal non-divisibility and is capable
of measuring it.

In order to compare with BHP measure, we also calculate the $g(t)$ defined by
Eq.~(\ref{eq:gt}). We get
\begin{equation}
g(t)  =  \frac{1}{2}\big[ |1-f(t) |+|f(t)|-1\big]
\end{equation}
with
\begin{equation}
  f(t) := -\omega\sin(2\omega t)\left(1-e^{-t}\right)+\cos^{2}(\omega t).
\end{equation}
The mediate dynamical maps $\mathcal{E}_{t+\epsilon,t}$ with infinitesimal
$\epsilon$ are not completely positive when $g(t)>0$. From
Fig.~\ref{fig:DB_Dtr}(b), we can see that the increasing of the Bures distance occurs in the regimes where $g(t)>0$, which coincides with the monotonicity of Bures distance under CPT maps.

\section{Conclusion\label{sec:Conclusion}}

In conclusion, we have shown that the measure for non-Markovianity
based on trace distance cannot reveal the infinitesimal
non-divisibility caused by the non-unital part of the dynamics. In
order to reflect effects of the non-unitality, we have constructed a
measure on the non-unital non-Markovianity, and also defined a
measure on the non-unitality, in the same spirit as BLP
non-Markovianity measure.

Like non-Markovianity, the non-unitality is another interesting
feature of the quantum dynamics. With the development of quantum
technologies, we need novel theoretical approaches for open quantum
systems. It is expected that some quantum information methods would
help us to understand some generic features of quantum dynamics. We
hope this work may draw attention to study more dynamical properties
from the informational perspective.

\begin{acknowledgments}
This work was supported by NFRPC through Grant No. 2012CB921602, the
NSFC through Grants No. 11025527 and No. 10935010 and National Research
Foundation and Ministry of Education, Singapore (Grant No. WBS: R-710-000-008-271). \end{acknowledgments}

\end{document}